 [2005/06/22 5.2/AAS markup document class]%
\documentclass[preprint]{aastex}%
\makeindex

\makeatletter 

\let\o@verbatim\verbatim
\def\verbatim{%
  \ifhmode\unskip\par\fi
  \ifx\@currsize\normalsize
     \small
  \fi
  \o@verbatim
}

\renewcommand \verbatim@font {%
  \normalfont \ttfamily
  \catcode`\<=\active
  \catcode`\>=\active
}

\RequirePackage{shortvrb}
\MakeShortVerb{\|}

\begingroup
  \catcode`\<=\active
  \catcode`\>=\active
  \gdef<{\@ifnextchar<\@lt\@meta}
  \gdef>{\@ifnextchar>\@gt\@gtr@err}
  \gdef\@meta#1>{\m{#1}}
  \gdef\@lt<{\char`\<}
  \gdef\@gt>{\char`\>}
\endgroup
\def\@gtr@err{%
   \ClassError{ltxguide}{%
      Isolated \protect>%
   }{%
      In this document class, \protect<...\protect>
      is used to indicate a parameter.\MessageBreak
      I've just found a \protect> on its own.
      Perhaps you meant to type \protect>\protect>?
   }%
}
\def\verbatim@nolig@list{\do\`\do\,\do\'\do\-}

\newcommand{\m}[1]{\mbox{$\langle$\it #1\/$\rangle$}}

\def\cmd#1{\cs{\expandafter\cmd@to@cs\string#1}}
\def\cmd@to@cs#1#2{\char\number`#2\relax}
\DeclareRobustCommand\cs[1]{\texttt{\char`\\#1}}
\def\GetFileInfo#1{%
  \def\filename{#1}%
  \def\@tempb##1 ##2 ##3\relax##4\relax{%
    \def\filedate{##1}%
    \def\fileversion{##2}%
    \def\fileinfo{##3}}%
  \edef\@tempa{\csname ver@#1\endcsname}%
  \expandafter\@tempb\@tempa\relax? ? \relax\relax}

\makeatother
\usepackage{spr-astr-addons}

\usepackage{amsfonts}
\usepackage{graphicx}
\usepackage{amsmath}
\usepackage{epsfig}
\usepackage{bm}

\RequirePackage{color}

\newcommand{\emaila}{e-mail:\,}

\expandafter\GetFileInfo\expandafter{\jobname.tex}%

\begin{document}

\tableofcontents


\title{Particles motion on topological Lifshitz black holes\\ in 3+1 dimensions}
\slugcomment{Not to appear in Nonlearned J., 45.}
\shorttitle{Particles motion on topological Lifshitz}
\shortauthors{Olivares, Rojas, V\'asquez \& Villanueva}

\author{ Marco Olivares}
\affil{Instituto de F\'{\i}sica,\\ Pontificia Universidad Cat\'{o}lica de Valpara\'{\i}so,\\
Av. Universidad 330, Curauma,\\ Valpara\'{\i}so, Chile.\\
\emaila{marco.olivaresrubilar@gmail.com}}
\author{Germ\'an Rojas }
\affil{North American College, Av. 21 de Mayo 833,\\ Arica, Chile.\\
\emaila{spanw17@gmail.com}}
\author{ Yerko V\'asquez }
\affil{Departamento de Ciencias F\'isicas,\\ Facultad de Ingenier\'ia,
Ciencias y Administraci\'on,\\ Universidad de La Frontera,\\ Avenida Francisco
Salazar 01145, Casilla 54-D,\\ Temuco, Chile.\\
\emaila{yvasquez@ufro.cl}}
\author{ J. R. Villanueva }
\affil{Departamento de F\'{\i}sica y Astronom\'{\i}a,\\ Universidad de Valpara%
\'{\i}so,\\ Gran Breta\~na 1111, Playa Ancha,\\ Valpara\'{\i}so, Chile.}
\affil{Centro de Astrof\'isica de Valpara\'iso, \\Gran Breta\~na 1111, Playa Ancha,\\ Valpara\'{\i}so, Chile.\\
\emaila{jose.villanuevalob@dfa.uv.cl}}

\maketitle

\begin{abstract}
 In the present paper we study the causal structure of a topological black hole
 presented by Mann R. B. JHEP \textbf{06}, 075 (2009) by mean the standard Lagrangian procedure, which
 allow us analyze qualitatively the behavior of test particles using the effective potential.
 Then, the geodesic motion of massive and massless particles is obtained analytically.
 We find that confined orbits are forbidden on this spacetime, however radial
 photons can escape to infinity in an infinite proper time but in a finite coordinate time, this
 correspond to an interesting and novel result.

\end{abstract}


\section{Introduction}

In the last years much attention have been focused in the application of the
AdS/CFT correspondence \citep{Maldacena:1997re} beyond high energy physics to
another areas of physics, where a more general class of spacetimes could be
the gravitational dual, for instance, to non relativistic scale invariant
theories of condensed matter physics. In \citep{Kachru:2008yh} has been
conjetured gravity duals of non relativistic Lifshitz -like fixed points
which describe multicritical points in certain magnetic materials and liquid
crystal, these curved duals are called Lifshitz spacetimes. An important
feature of these spacetimes is its invariance under an anisotropic scale
transformation. There are many theories with anisotropic invariant scale of
interest in studying such critical points instead of the scale invariance
which arises in the conformal group, particularly in the studies of critical
exponent theory and phase transitions. Systems with these features have
appears in the description of strongly correlated electrons of strange
metals \citep{Hartnoll:2009ns}. The scale invariance is expressed as $%
t\rightarrow \lambda ^{z}t$, $x\rightarrow \lambda x$, where $z\neq 1$ is
the critical exponent which measure the degree of anisotropy between spatial
and temporal scalings.

On the gravitational side, $D$-dimensional spacetimes that exhibit these
symmetries are described by the Lifshitz metrics

\begin{equation}
ds^{2}=-\frac{r^{2z}}{l^{2z}}dt^{2}+\frac{l^{2}}{r^{2}}dr^{2}+\frac{r^{2}}{%
l^{2}}d\vec{x}^{2},  \label{metric}
\end{equation}
where $\vec{x}$ represents a $D-2$ dimensional spatial vector and $l$
denotes the length scale in the geometry. If $z=1$, the spacetime is the
usual anti-de Sitter metric in Poincar\'{e} coordinates and have the larger
symmetry $SO(D-1,2)$. All curvature invariants of metric (\ref{metric}) are
constant and these spacetimes have a null curvature singularity at $%
r\rightarrow 0$ for $z\neq 1$, this can be seen by computing the tidal
forces between infalling particles. This singularity is reached in finite
proper time by infalling observers so the spacetime is geodesically
incomplete \citep{Horowitz:2011gh}. A natural extension of the above
spacetime is consider black hole solutions in this background whose
asymptotic behavior is given by (\ref{metric}), these are called
asymptotically Lifshitz black holes \citep{Balasubramanian:2009rx}-\citep%
{{Dehghani:2010kd}} .

We are interesting in analyze the geodesic structure of a class of
asymptotically Lifshitz black hole recently found in the literature,
performing a study of freely moving of test particles and photons. The
geodesic equations of motions are a set of ordinary differential equations
describing the evolution of the coordinates of these test particles as a
function of an affine parameter. If the spacetime have symmetries the
conserved quantities associate to these symmetries simplify the analysis of
the equations. The geodesic structure of the Schwarzschild,
Reissner-Nordstr\"{o}m and Kerr black holes were studied by Chandrasekhar \citep%
{Chandrasekar}. The geodesic study in a Schwarzschild spacetime background
was the key to understand various astrophysical phenomena, as planetary
motion, gravitational lensing and radar delay among others. Besides of
astrophysical objects it is important to study, motivated by the AdS/CFT
correspondence \citep{Maldacena:1997re} and its generalizations, the geodesic
structure of black holes with an asymptotic behavior different than the flat
case, furthermore these black holes provide a theoretical laboratory for
understanding interesting features of black holes physics. In this regard,
geodesics around the Schwarzschild anti-de Sitter black hole was studied in
\citep{Kraniotis}-\citep{Olivares}-\citep{HackmanA}-\citep{HackmannB},
and the Schwarzschild de Sitter case in \citep{Hellaby}-\citep{Calvani}-\citep{Podolsky}.
The motion of uncharged particles in Reissner-Nordstr\"{o}m
black hole with a non-zero cosmological constant has been studied in \citep{Z}-\citep{cosv},
while the study of charged particles can be found in \citep{osvl}.
Orbits in Kerr and Kerr (anti) de Sitter are calculated
in \citep{Kraniotis2}.
Study of geodesic motion of test particles in higher dimensional
Schwarzschild, Schwarzschild anti-de Sitter, Reissner-Nordstr\"{o}m and Reissner-Nordstr\"{o}m
anti-de Sitter spacetimes can be found in references \citep{Kunz}-\citep{Gibbons},
where complete solutions and a classification of the
possible orbits in these geometries in term of elliptic functions have
been obtained.

In this article we will study the geodesic structure of the topological
Lifshitz black hole in 3+1 dimensions with critical exponent $z=2$ \citep%
{mann}. We will perform an analysis of particle motions by means of an
effective potential, we will provide graphics of the orbits and classify
different kinds of motion for particles by the values of its angular
momentum.

This paper is organized as follows. In Sec. 2, we briefly review some features of topological
Lifshitz black hole. In Sec. 3 we obtain the radial
equation of motion for geodesics with an effective potential. Therefore we
study null and time-like geodesics. We conclude
in Sec. 4 with some final remarks.

\section{Topological Lifshitz Black Hole}

\begin{figure}[h]
\begin{center}
\includegraphics[width=75mm]{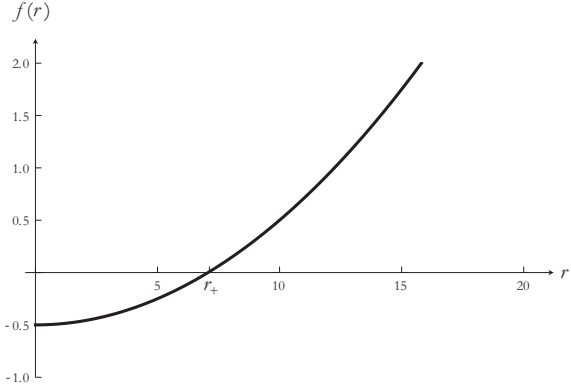}
\end{center}
\caption{Plot of the lapse function as a function of the radial coordinate, $%
r$, with $\ell =10$ ($r_+=5\sqrt{2}$).}
\label{f1}
\end{figure}

Four dimensional topological Lifshitz black hole was presented in \citep{mann}
where the author obtained a black hole solution with critical exponent $z=2$%
, which posses an event horizon if the curvature of the transverse spatial
sections is negative $k=-1$. These kinds of black holes are described by the
following metric,
\begin{equation}
ds^{2}=-\frac{r^{2}f(r)}{\ell ^{2}}dt^{2}+\frac{dr^{2}}{f(r)}+r^{2}(d\theta
^{2}+\text{sinh}^{2}\theta d\phi ^{2}),  \label{e1}
\end{equation}%
where the lapse function $f(r)$ is given by%
\begin{equation}
f(r)=\frac{r^{2}}{\ell ^{2}}-\frac{1}{2},  \label{e2}
\end{equation}%
and the coordinates are defined in the intervals $-\infty <t<\infty $, $r>0$%
, $0\leq \phi \leq 2\pi $, and $0\leq \theta \leq 2\pi $. In FIG. \ref{f1}
we show a graph of the lapse function for topological Lifshitz black holes.

The event horizon is located at $r_{+}=\ell /\sqrt{2}$, and corresponds to a
coordinate singularity, which can be seen from the Ricci scalar,
\begin{equation}
R=\frac{1}{r^{2}}-\frac{11}{r_+^{2}},  \label{e3}
\end{equation}%
such that,
\begin{equation}
R_{\,r=r_{+}}=-\frac{10}{r_{+}^{2}}.  \label{e4}
\end{equation}

The principal quadratic invariant of the Ricci tensor and the Kretschmann scalar
are given by the following expressions

\begin{equation}\label{r1}
R_{\mu \nu }R^{\mu \nu }=\frac{33r^{4}-8\,r_+^{2}r^{2}+r_+^{4}}{r_+^{4}\,r^{4}},
\end{equation}
and
\begin{equation}\label{r2}
R_{\mu \nu \rho \sigma }R^{\mu \nu \rho \sigma }=\frac{3\left(
r_+^{4}-2\,r_+^{2}r^{2}+9r^{4}\right) }{r_+^{4}\,r^{4}},
\end{equation}
respectively. These invariants tell us that the physical singularity is located at $r=0$.

In \citep{Pablo} a study of the quasinormal modes and the absorption cross
section for this black hole was performed, and was shown that it is stable
under scalar field perturbations.

Finally, it's interesting to notice that metric (\ref{e1}) looks analogue to
a zero-mass topological AdS black hole \citep{mann2}.

\section{Geodesic Structure}

In order to compute the geodesic structure of the topological Lifshitz black
hole, we will use the standard Lagrangian procedure \citep{Olivares}, and thus,
we write the Lagrangian associated to the metric (\ref{e1}), resulting
\begin{equation}
2\mathcal{L}=-\frac{r^{2}f(r)}{2\, r_+^{2}}\dot{t}^{2}+\frac{\dot{r}^{2}}{f(r)}%
+r^{2}(\dot{\theta}^{2}+\text{sinh}^{2}\theta \,\dot{\phi}^{2})=-m,
\label{e5}
\end{equation}
where the dot corresponds to derivative with respect to an affine parameter
along the geodesic, and, by normalization, $m=1$ for massive particles and $%
m=0$ for massless particles.

Since ($t,\phi $) are cyclic coordinates, their conjugate momenta, $\Pi _{q}$%
, are conserved. In our case we obtain,
\begin{equation}
\Pi _{\phi }=r^{2}\,\dot{\phi}\,\text{sinh}^{2}\theta =L,  \label{e6}
\end{equation}%
and,
\begin{equation}
\Pi _{t}=-\frac{r^{2}}{2\,r_+^{2}}f(r)\dot{t}=-\sqrt{E}.  \label{e7}
\end{equation}%
Here $L$ is the angular momentum, but the constant of motion $E$ cannot be
associated globally to the energy because this spacetime is not
asymptotically flat. On the other hand, eq. (\ref{e6}) implies that the
motion is performed in an invariant plane, which, for simplicity, we choose
to be the plane defined by $\theta =\theta _{0}$, such that, $\text{sinh}%
\,\theta _{0}=1$. In this way, we rewrite eq. (\ref{e6}) as
\begin{equation}
r^{2}\,\dot{\phi}=L.  \label{e8}
\end{equation}

Then, introducing eqs. (\ref{e7}) and (\ref{e8}) into eq. (\ref{e5}), we
obtain the radial equation of motion,
\begin{equation}
\dot{r}^{2}=2\,\frac{\,r_+^2}{r^{2}}\,\left[ E-V(r)\right] ,  \label{e9}
\end{equation}%
where $V(r)$ is the effective potential given by
\begin{equation}
V(r)=\frac{1}{4}\,\frac{r^{2}}{r_+^{2}}\left(\frac{r^{2}}{r_+^{2}}-1\right)\,\left( m+\frac{L^{2}}{r^{2}}\right).
\label{e10}
\end{equation}%
In the next sections, based in this effective potential, we will study all
possible motion for massive and massless particles.

\subsection{Null Geodesics}

Null geodesic corresponds to a photon trajectory, i. e., $m=0$, so, the
effective potential, (\ref{e10}), takes the form,
\begin{equation}
V_{n}(r)=\frac{1}{4}\,\frac{L^{2}}{r_+^{2}}\left(\frac{r^{2}}{r_+^{2}}-1\right).  \label{e11}
\end{equation}

\subsubsection{Radial Motion}

Radial photons are characterized by vanishing angular momentum ($L=0$),
which implies that effective potential vanishes too,
\begin{equation}
V_{nr}(r)=0,  \label{e12}
\end{equation}%
and thus, photons have the possibility of not falling into the black hole
and escape indefinitely towards spatial infinity. Therefore, eq. (\ref{e9})
becomes
\begin{equation}
\dot{r}^{2}=2E\,\frac{r_+^{2}}{r^{2}},  \label{e13}
\end{equation}%
so, an elementary integration yields to
\begin{equation}
\tau (r)=\pm\frac{1}{2} \frac{R_{0}^{2}}{ \sqrt{2E}\,r_+}\left[ \left( \frac{r}{R_{0}}%
\right) ^{2}-1\right] ,  \label{e14}
\end{equation}%
where $R_{0}$ corresponds to the radial distance in which $\tau =0$.

Next, making use of the identity
\begin{equation}
\frac{dr}{d\tau }=\left( \frac{dr}{dt}\right) \dot{t},  \label{e15}
\end{equation}%
together with eqs. (\ref{e7}) and (\ref{e13}), we obtain the quadrature
\begin{equation}
\frac{dr}{dt}=\pm\frac{1}{2\sqrt{2}} \frac{r}{r_+}\left(\frac{r^{2}}{r_+^{2}}-1\right).  \label{e16}
\end{equation}%
Therefore, integrating this last equation from $R_{0}$ ($t(r=R_{0})=0$) to $%
r $, we obtain 
\begin{equation}
t(r)=\pm \sqrt{2}\,r_+ \left[ \ln \left\vert \frac{r^2-r_{+}^2}{R_{0}^2-r_{+}^2}\right\vert
-\ln \left(
\frac{r}{R_{0}}\right)^2 \right] .  \label{e17}
\end{equation}%
\begin{figure}[h]
\begin{center}
\includegraphics[width=70mm]{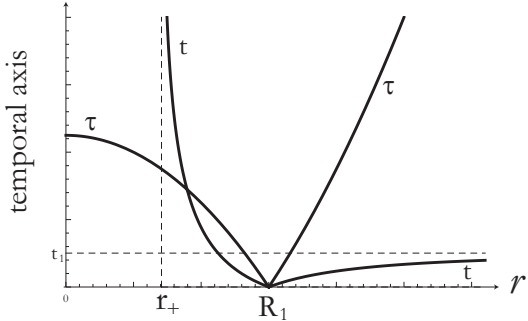}
\end{center}
\caption{Plot of the proper time, $\protect\tau (r)$, and coordinate time, $%
t(r)$, as a function of the radial coordinate $r$ with $\ell =10$.}
\label{f2}
\end{figure}
In FIG. \ref{f2} we show the proper time $\tau $ and the coordinate time $t$
as a function of the radial coordinate $r$. As in Schwarzschild spacetime,
photons fall into the black hole in a finite proper time, however an
external observer see that photons fall asymtotically to the black hole.
Moreover, the topological Lifshitz black hole admits radial photons come to
infinity, and, in the proper system they come to infinity in infinity proper
time, but an external observer see that they come to infinity in a finite
coordinate time, $t_{1}$, given by
\begin{equation}
t_{1}=\lim\limits_{r\rightarrow \infty }t(r)=\sqrt{2}\,r_+ \,\ln \left\vert \frac{%
R_{0}^{2}}{R_{0}^{2}-r_{+}^{2}}\right\vert .  \label{e18}
\end{equation}%
This situation is unique with respect to the Einstein's spacetimes.

\subsubsection{Angular Motion}

\begin{figure}[!h]
\begin{center}
\includegraphics[width=70mm]{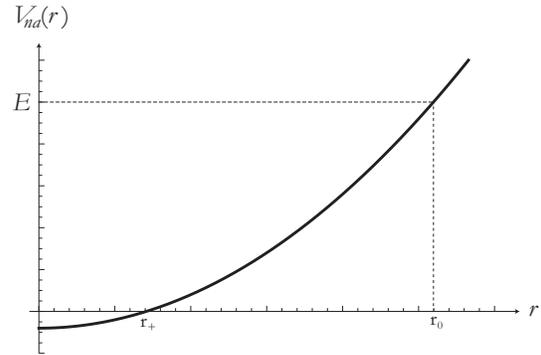}
\end{center}
\caption{Effective potential for photons in angular motion, $V_{na}(r)$,
with $r_+=5\sqrt{2}$ ($\ell=10$) and $L=4$.}
\label{f3}
\end{figure}

In this case photons posses angular momentum $L>0$, thus, its effective
potential is
\begin{equation}
V_{na}(r)=\frac{1}{4}\,\frac{L^{2}}{r_+^{2}}\left(\frac{r^{2}}{r_+^{2}}-1\right),  \label{e181}
\end{equation}%
which is shown in FIG. \ref{f3}.

Next, using the identity
\begin{equation}
\dot{r}=\left( \frac{dr}{d\phi }\right) \dot{\phi},  \label{e182}
\end{equation}%
\begin{figure}[h]
\begin{center}
\includegraphics[width=60mm]{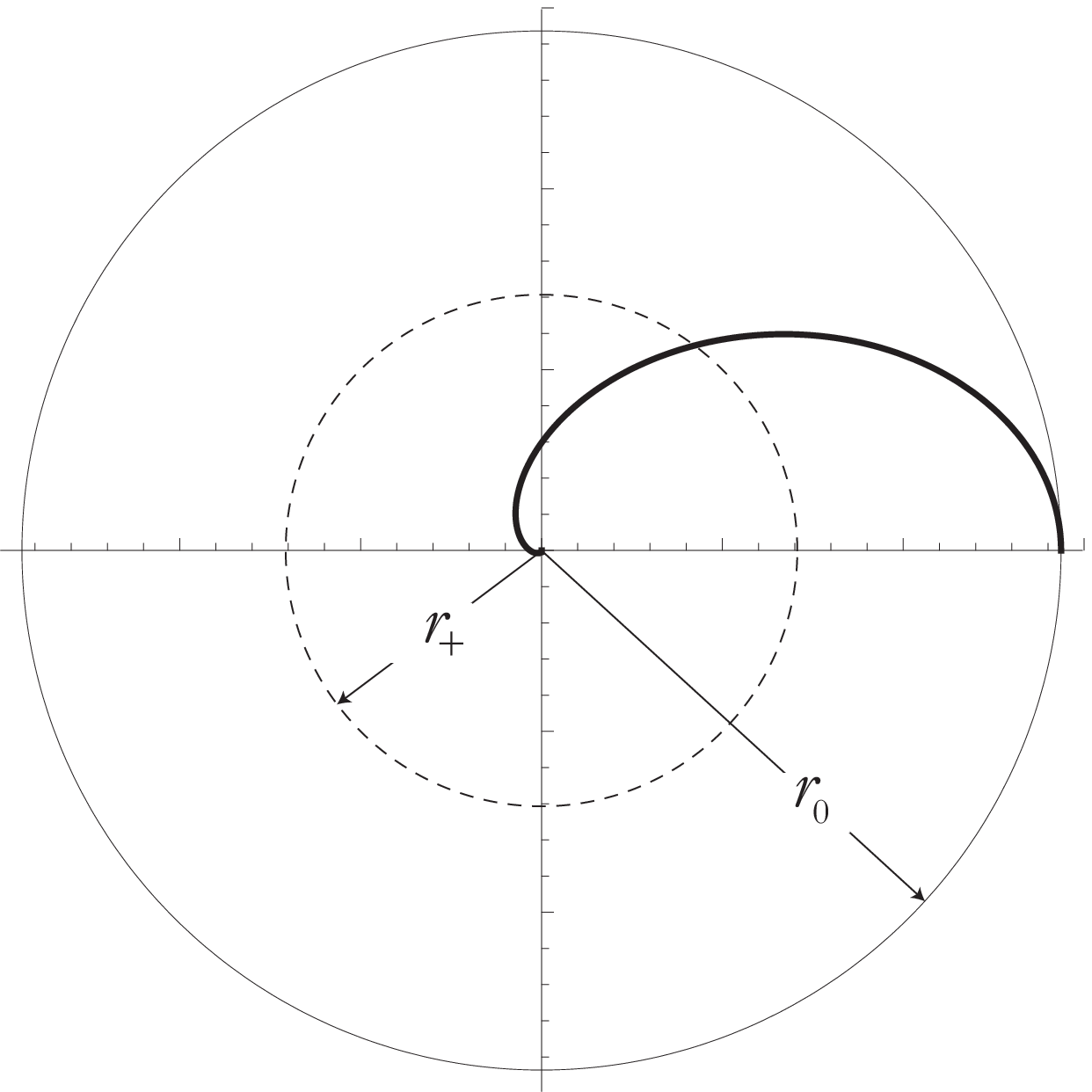}
\end{center}
\caption{Polar trajectory (\protect\ref{e22}) with $r_+=5\sqrt{2}$ ($\ell=10$) and $L=4$.}
\label{f4}
\end{figure}
together with the motion integral (\ref{e6}), eq. (\ref{e8}) can be written
as
\begin{equation}
\left( \pm \frac{dr}{d\phi }\right) ^{2}=2\,r_+^{2}\,r^{2}\left( \frac{1}{%
\mathfrak{D}^{2}}-\frac{1}{4}\frac{r^{2}}{r_+^{4}}\right) ,  \label{e19}
\end{equation}%
%
where $\mathfrak{D}=\left( \frac{1}{b^{2}}+\frac{1}{4}\frac{1}{r_+^{2}}\right)
^{-1/2}$ is the \textit{anomalous impact parameter}, and $b=\sqrt{L^{2}/E}$
corresponds to the usual impact parameter. Now, making the change of
variable $x=\frac{r}{\sqrt{2}\,r_+ }$, we obtain
\begin{equation}
\phi (x)=\int_{x_{0}}^{x}\frac{-dx}{x\cdot \sqrt{\mathcal{P}(x)}},
\label{e20}
\end{equation}%
%
where $\mathcal{P}(x)$ is a second order polynomial given by $\mathcal{P}%
(x)=x_{0}^{2}-x^{2}$, with $x_{0}=\frac{r_{0}}{\sqrt{2}\,r_+} $, and $r_{0}$ corresponds to
the value of $r$ when $t=\tau =0$. Integrating eq. (\ref{e20}), we obtain
\begin{equation}
\phi (x)=\frac{1}{x_{0}}\text{arccosh}\left( \frac{x_{0}}{x}\right) ,
\label{e21}
\end{equation}
%
%
therefore, the polar trajectory is given by
\begin{equation}
r(\phi )=r_{0}\,\text{sech}\left( \frac{r_{0}}{\sqrt{2}\,r_+}\phi \right),
\label{e22}
\end{equation}%
which is shown in FIG. \ref{f4}.

\subsection{Time-Like Geodesics}

In this section we will consider the motion of massive particles on the
background of the topological Lifshitz black hole. In this case we have $m=1$%
, therefore, the effective potential (\ref{e9}) acquires the form%
\begin{equation}
V_{t}(r)=\frac{1}{4}\,\frac{r^{2}}{r_+^{2}}\left(\frac{r^{2}}{r_+^{2}}-1\right)\,\left( 1+\frac{L^{2}}{r^{2}}\right).
\label{e22.1}
\end{equation}%
%

\subsubsection{Radial Motion}

\begin{figure}[h]
\begin{center}
\includegraphics[width=70mm]{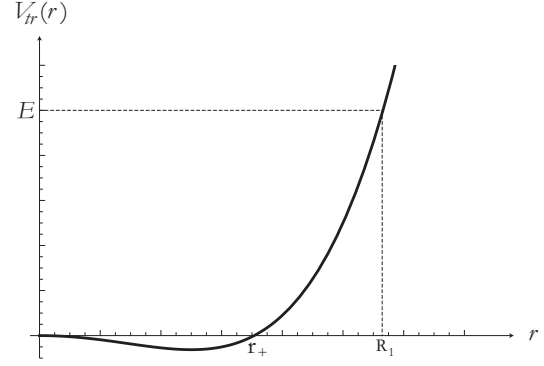}
\end{center}
\caption{Effective potential for massive radial particles, $V_{tr}(r)$, with
$r_+=5\sqrt{2}$ ($\ell=10$) and $L=4$.}
\label{f5}
\end{figure}

For this situation the particles have zero angular momentum $L=0$, and the
effective potential (\ref{e22.1}) simplifies to
\begin{equation}
V_{tr}(r)=\frac{1}{4}\,\frac{r^{2}}{r_+^{2}}\left(\frac{r^{2}}{r_+^{2}}-1\right),  \label{e23}
\end{equation}%
this potential is shown in FIG. \ref{f5}.

\begin{figure}[h]
\begin{center}
\includegraphics[width=65mm]{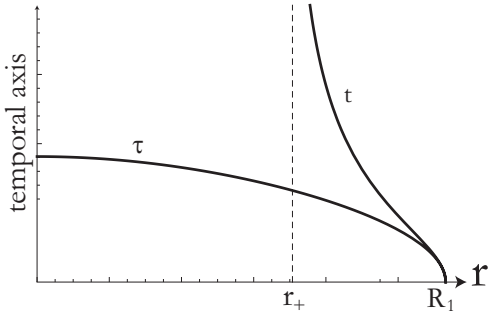}
\end{center}
\caption{Proper and coordinate time for radial massive particles as a
function of radial coordinate $r$ with $r_+=5\sqrt{2}$ ($\ell=10$) and $L=4$.}
\label{f6}
\end{figure}

In this way, eqs. (\ref{e8}) and (\ref{e23}) lead to the quadrature
\begin{equation}
\left( \frac{dr}{d\tau }\right) ^{2}=2\frac{r_+^{2}}{r^{2}}\left[ E-
\frac{1}{4}\,\frac{r^{2}}{r_+^{2}}\left(\frac{r^{2}}{r_+^{2}}-1\right) \right] ,
\label{e24}
\end{equation}%
so that making the change of variable $r=r_+\,\sqrt{2y} $, eq. (\ref{e24})
becomes

\begin{equation}
\tau (y)=-\frac{r_+}{\sqrt{2}}\,\int_{y_{0}}^{y}\frac{dy}{\sqrt{E+\frac{y}{2}-y^{2}}%
},  \label{e26}
\end{equation}%
where we assume that $\tau(y_0)\equiv \tau_0=0$.
So, integrating eq. (\ref{e26}) we find
\begin{equation}
\sqrt{2}\,\frac{\tau (r)}{ r_+}=\frac{\pi}{2}-\arcsin \left( \frac{2\,r^2-r_+^2}{2\,R_1^2-r_+^2}\right),
\label{e27}
\end{equation}%
%
and therefore, we obtain the final expression
\begin{equation}
r(\tau )=\frac{r_+}{\sqrt{2}}\sqrt{1+\left(2\frac{R_1^2}{r_+^2}-1\right)\sin \left( \frac{\pi }{2}-\sqrt{2}%
\frac{\tau }{r_+}\right) }.  \label{e29}
\end{equation}%
On the other side, using (\ref{e7}) and (\ref{e15})
into eq. (\ref{e24}), and then integrating
the quadrature, we obtain
\begin{eqnarray}\nonumber
  \frac{t(r)}{\sqrt{2}\,r_+}&=&\textrm{arccosh}\left[\frac{8E+\frac{r^2}{r_+^2}-1}{\left(\frac{r^2}{r_+^2}-1\right)\left(2\frac{R_1^2}{r_+^2}-1\right)}\right]+\\ \label{e291}
   &-&\textrm{arccosh}\left[\frac{8E+\frac{r^2}{r_+^2}}{\frac{r^2}{r_+^2}\left(2\frac{R_1^2}{r_+^2}-1\right)}\right] +\varpi_{t},
\end{eqnarray}
where
\begin{eqnarray}\nonumber
\varpi_t&=& \textrm{arccosh}\left[\frac{8E+\frac{R_1^2}{r_+^2}}{\frac{R_1^2}{r_+^2}\left(2\frac{R_1^2}{r_+^2}-1\right)}\right]+\\\label{e292}
   &-&\textrm{arccosh}\left[\frac{8E+\frac{R_1^2}{r_+^2}-1}{\left(\frac{R_1^2}{r_+^2}-1\right)\left(2\frac{R_1^2}{r_+^2}-1\right)}\right],
\end{eqnarray}
and the turning point, $R_1$, is given by
\begin{equation}\label{e293}
  R_1=\sqrt{\frac{1+\sqrt{1+16E}}{2}}\,r_+.
\end{equation}
%
%
FIG. \ref{f6} shows a graph of eqs. (\ref{e29}) and (\ref{e291}).


\subsubsection{Angular motion}

\begin{figure}[h]
\begin{center}
\includegraphics[width=75mm]{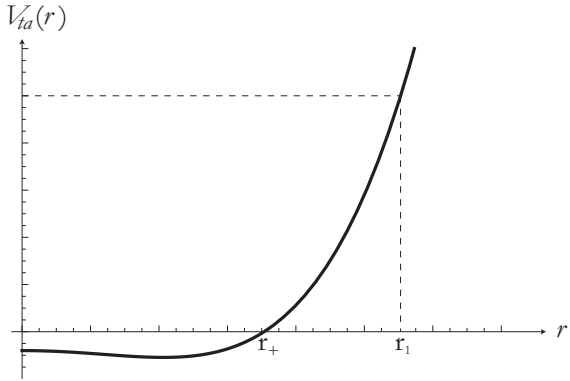}
\end{center}
\caption{Effective potential of angular massive particles, $V_{ta}(r)$, with
$r_+=5\sqrt{2}$ ($\ell=10$) and $L=4$.}
\label{f7}
\end{figure}

Particles with angular motion are characterized by $L>0$. The effective
potential in this case is given by
\begin{equation}
V_{ta}(r)=\frac{1}{4}\,\frac{r^{2}}{r_+^{2}}\left(\frac{r^{2}}{r_+^{2}}-1\right)\,\left( 1+\frac{L^{2}}{r^{2}}\right),
\label{e30}
\end{equation}%
which is depicted in FIG. \ref{f7}.

\begin{figure}[!h]
\begin{center}
\includegraphics[width=60mm]{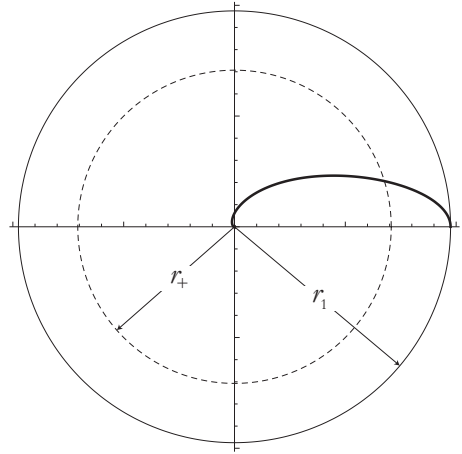}
\end{center}
\caption{Polar trajectory (\protect\ref{e34}) with $r_+=5\sqrt{2}$ ($\ell=10$)
and $L=4$.}
\label{f8}
\end{figure}

Using eqs. (\ref{e6}), (\ref{e18}) and (\ref{e30}) we get
\begin{equation}
\left( \frac{dr}{d\phi }\right) ^{2}=2\frac{r_+^{2}\,r^{2}}{L^{2}}\left[ E-%
\frac{1}{4}\,\frac{r^{2}}{r_+^{2}}\left(\frac{r^{2}}{r_+^{2}}-1\right)\,\left( 1+\frac{L^{2}}{r^{2}}\right) \right] ,  \label{e31}
\end{equation}%
%
%
%
%
%
%
%
and, after the change of variable $r=r_+\,\sqrt{2y}$ in the above equation, we
obtain the following expression

\begin{equation}
\phi (x)=-\frac{1}{4}\,\frac{L^2}{r_+^2}\, \int_{y_{0}}^{y}\frac{dy}{y\sqrt{\mu -\sigma
y-y^{2}}},  \label{e32}
\end{equation}%
where
\begin{equation}\label{e33}
  \mu=E+\frac{1}{8}\,\frac{L^4}{r_+^4},\quad \textrm{and}\quad
  \sigma=\frac{1}{4}\,\frac{L^4}{r_+^4}-\frac{1}{2}.
\end{equation}
%
%
%
%
%
Making the integration and returning to the original variables we obtain an
analytical expression for $r(\phi )$, the polar form of the orbit
\begin{equation}
r(\phi )=r_1 \sqrt{\frac{1+\Omega}{1 +\Omega \cosh (\alpha
_{0}\phi )}},  \label{e34}
\end{equation}%
%
%
where the turning point is given explicitly by
\begin{equation}\label{e35}
  r_1=\frac{2\sqrt{\mu/\sigma}\,r_+}{\sqrt{1+\Omega}},
\end{equation}
and the constants $\Omega $ and $\alpha _{0}$ are defined as
\begin{equation}\label{e36}
  \Omega=\sqrt{1+4\frac{\mu}{\sigma^2}},\quad \textrm{and}\quad
  \alpha_0=\frac{4\sqrt{\mu}\,r_+^2}{L^2},
\end{equation}
respectively. FIG. \ref{f8} shows the polar trajectory of massive
particles (\ref{e34}).

\section{Summary}
In this work we have studied the causal structure of topological Lifshitz
black hole using the standard Lagrangian procedure. We analyzed the radial
and angular motion of massless and massive test particles on this
background. Because the metric coefficients doesn't depends on the temporal and
azimuthal coordinates, it allows us to find two
integrals of motion, and reduce the system of equations to one radial
equation of motion with an effective potential. We have obtained analytical
expressions for the proper time and coordinate time as a function of the
radial coordinate as well as analytical expressions for the polar form of
the orbits in all cases considered. For radial photons we obtain that the
solutions show a similar behavior to what occurs in the Schwarzschild black
hole, in the sense that in the proper time framework, these radial massless
particles reach the event horizon and cross it in a finite proper time,
however in the coordinate time framework, the external observer sees that
takes an infinite time the particles to reach the horizon. Besides, on other
hand, topological Lifshitz black hole admits radial photons come to infinity
in a finite coordinate time as seen by an external observer, these results
are depicted in FIG 2. For photons with angular momentum, the effective
potential is unbounded and increases with the radial distance acting as an
attractive force, as show FIG. 3. Therefore, non radial photons always will
reach the event horizon, this can be seen additionally in FIG. 4 where the
polar trajectory of the photon is illustrated. For massive particles the
effective potential have a qualitatively similar behavior for both radial
and non-radial particles. Analogous to the above case of non-radial photons,
these potentials increase with the radial distance, which means that the
particles are confined and all them will reach the event horizon. Therefore
the space-time doesn`t admit bounded orbits. In FIG 8 we show the orbit of a
generic non-radial massive particle.

\begin{acknowledgments}
Y. V. is supported by FONDECYT grant 11121148. M. O.
thanks to PUCV.
\end{acknowledgments}


\end{document}